# Disconnected, fragmented, or united? A trans-disciplinary review of network science

César A. Hidalgo | Associate Professor | The MIT Media Lab
hidalgo@mit.edu

**Abstract:**

During decades the study of networks has been divided between the efforts of social scientists and natural scientists, two groups of scholars who often do not see eye to eye. In this review I present an effort to mutually translate the work conducted by scholars from both of these academic fronts hoping to continue to unify what has become a diverging body of literature. I argue that social and natural scientists fail to see eye to eye because they have diverging academic goals. Social scientists focus on explaining how context specific social and economic mechanisms drive the structure of networks and on how networks shape social and economic outcomes. By contrast, natural scientists focus primarily on modeling network characteristics that are independent of context, since their focus is to identify universal characteristics of systems instead of context specific mechanisms. In the following pages I discuss the differences between both of these literatures by summarizing the parallel theories advanced to explain link formation and the applications used by scholars in each field to justify their approach to network science. I conclude by providing an outlook on how these literatures can be further unified.

**Introduction: Born fragmented**

> *"Science must, over the next 50 years, learn to deal with these problems of organized complexity."—Warren Weaver, 1948*

How science evolves? And how is scientific progress tied to improvements in mathematics? In 1948 Warren Weaver, the director of the Rockefeller Foundation's division of natural sciences, published an essay hoping to answer these questions.

His now classic paper: *Science and Complexity* (Weaver 1948); explained the three eras that according to him defined the history of science. These were the era of *simplicity*, *disorganized* *complexity*, and *organized complexity*. In the eyes of Weaver what separated these three eras was the development of mathematical tools allowing scholars to describe systems of increasing complexity.

The first era, that of *problems of simplicity*, focused on systems that could be described using trajectories and surfaces. These are the systems that could be modeled using the calculus developed by Newton and Leibniz. Of course, there are many *problems of simplicity* that are mathematically complicated, but in the eyes of Weaver these mathematical complications were not the same as complexity, since complexity could only emerge in systems populated by many interacting components. These are systems that evolve, adapt, and beget diversity in ways that cannot be well-described using calculus, so for science to continue its progress, a new math needed to emerge.

That new math was statistics and probability, which allowed scholars to focus on a new class of problems: problems of *disorganized complexity*. Problems of disorganized complexity are problems that can be described using averages and distributions, and that do not depend on the identity of the elements involved in a system, or their precise patterns of interactions. A classic example of a problem of disorganized complexity is the statistical mechanics of Ludwig Boltzmann, James-Clerk Maxwell, and Willard Gibbs, which focuses on the properties of gases. Here, each molecule inside a gas can be considered to be the same. These problems also involve the mathematical reformulation of Darwin's theory evolution advanced by Karl Pearson, Sewall Wright, Jack Haldane, and Ronald Fisher, which focused on the coarse patterns that generate from combining variation and selection. But the probability and statistics methods that helped advanced our understanding of systems of disorganized complexity still had limitations, as it could not account for the complex patterns begot in the intimacy of society and life.

So in the midst of the twentieth century Weaver saw the dawn of a new era: the era of *organized complexity*. This was a new science focused on problems where the identity of the elements involved in a system, and their patterns of interactions, could no longer be ignored. This involved the study of biological, social, and economic systems. According to Weaver, to make progress in the era of organized complexity, a new math needed to emerge.

Since Weaver published his seminal paper scholars have improved our understanding of systems of organized complexity. Part of this progress involves the development of the science of networks, which is a clear response to Weaver's request. Networks are mathematical objects that help us keep track of the identity of the elements involved in a system and their patterns of interactions, making networks the ideal structures to describe problems of organized complexity. Of course, networks are no panacea, or represent a complete toolbox, but together with the tools of calculus, probability, and statistics; they provide us with a more comprehensive toolbox that we can use to describe systems of organized complexity and test hypothesis about how these work.

But the study of systems of organized complexity did not grow radially from Weaver's seminal paper, or from a single stream of literature. Instead, it grew in patches, in independent and often unconnected parts of academia. Unlike other academic efforts, which usually grow from a single academic source, the science of organized complexity was born fragmented, with pioneers in many different fields. Soon after Weaver's paper, biologists like Francois Jacob (Jacob and Monod 1961), (Jacob et al. 1963) and Stuart Kaufmann (Kauffman 1969), developed the idea of regulatory networks. Mathematicians like Paul Erdos and Alfred Renyi, advanced graph theory (Erdős and Rényi 1960) while Benoit Mandelbrot worked on Fractals (Mandelbrot and Van Ness 1968), (Mandelbrot 1982). Economists like Thomas Schelling (Schelling 1960) and Wasily Leontief (Leontief 1936), (Leontief 1936), respectively explored self-organization and input-output networks. Sociologists, like Harrison White (Lorrain and White 1971) and Mark Granovetter (Granovetter 1985), explored social networks, while psychologists like Stanley Milgram (Travers and Milgram 1969) explored the now famous small world problem. The science of organized complexity emerged in the second half of the twentieth century, just as Weaver predicted, but it emerged in parallel efforts that are not easy to reconcile.

My goal in this paper is not to reconcile these streams of literature—that would be too ambitious—but to create a narrative that translates the value of the research conducted in one stream of literature to scholars from other streams. To achieve this goal, however, I will need to make some coarse simplifications. For simplicity, I will divide network science into two main streams, the streams advanced by social scientists, and pioneered by sociologists, political scientists, and economists—who of course, have important differences among them—and the stream of literature advanced by the natural scientists, which is dominated by scholars trained as physicists, computer scientists, mathematicians, and biologists, from geneticists to ecologists. Certainly, there are important differences within each of these groups and subgroups. In the context of the social sciences, economists tend to focus more on the creation of formal models built on ideas of utility maximization than sociologists, and are also, more obsessed with methods for causal identification than sociologists, even though sociologists are no strangers to causal inference. Computer scientists are also quite different from physicists, since they tend to focus more, for instance, on the optimization of algorithms than on the universality of distributions. But nevertheless, these within group differences can often be small compared to the differences observed between groups, so I will nevertheless take a first pass at painting this picture using a thick brush. I apologize to those that will take this simplification with outrage.

Also, some people may argue that the division between these sciences is no longer present, since there has been an increase in multidisciplinary efforts that transcend traditional boundaries. There are now, for instance, new degree programs on network science that have hired scholars from multiple disciplines (Lazer et al. 2009). At the same time, we should not overgeneralize from a few examples. These examples, while encouraging, may not be representative of all academic departments, and could be in fact, seen as the emergence of yet another group. So with the danger of oversimplifying I will focus on dividing academia into a few coarse groups for two reasons. First, I will focus in these larger groups because reviews that transcend the boundary between the social and natural sciences are rare, but I believe them to be valuable. One such review is Borgatti et al. (2009), which compares the network science of natural and social scientists arriving to a similar conclusion to the one I arrived. Second, I believe these diverging bodies of literature are in desperate need of mutual understanding, and to achieve that understanding, we need to help translate the research goals and intentions of one group of researchers to the language of the other (or at least, into a simple language that everyone can understand). Of course, the breadth of the effort implies that I am destined to fall short, and make a review that is both, narrow and incomplete. Also, for those who are experts in a particular stream of literature, parts of this review will seem dated, since I am not focusing on what is more recent, but on the historical trajectories of the ideas advanced by scholars in each of these streams. For a comprehensive summary of the literature advanced by a particular branch of the literature I recommend readers to look at reviews that focus on more narrow subjects. The purpose of this review, therefore, is not to summarize all of the streams of literature that discuss networks[1], but to pick a few illustrative examples that can help translate the goals advanced by scholars working on different corners of what is a vast intellectual space. I hope this exercise is useful for the growing community of scholars working on networks, and also, that it contribute to the educational efforts needed to establish the study of networks as a field.

**Links and Link Formation**

Links are the essence of networks. So I will start this review by comparing the mechanisms used by natural and social scientists to explain link formation. Before I describe these mechanisms, however, I will note that even the notion of what is considered a link can be different for scholars in both streams of literature.

Social scientists' idea of links—or ties—often incorporates information on the context of social interactions and the type of support that flows through that interaction. For instance, social scientists make strong differences between friendship ties, co-working ties, and family ties, because different types of links provide different forms of support and affect the dynamics of different aspects of society. Even more, among family ties, social scientists will often differentiate between the ties connecting parents to their offspring and to each other, since these are relationships ruled by a different set of norms and expectations. Also, in the context of the literature on social capital, social scientists interpret ties as the embodiment of trust (Granovetter 1985), (Fukuyama 1995), (Coleman 1988). So in the social science literature, and in particular in the literature advanced by sociologists and political scientists, links are not simply a recollection of instances of communication, but social relationships that are meaningful only as long as the individuals involved in them trust and support each other in specific ways.

Natural scientists' definition of links, however, has been more abstract and driven by the availability of data. Their implicit definition of connections involves recorded acts of communications that are independent of social context (i.e. a phone call or email), a technological link (a URL in a webpage or a physical link in the internet), or collaboration in a creative process (paper co-authorships or sharing credits in a movie). This contrasts with the definitions preferred by social scientists, where the type of relationships is considered important. For instance, a co-authorship link is not the same if it is between two students, or between a student and his or her advisor. When connecting the people that acted in the same movie, natural scientists do not differentiate between people in leading or supporting roles. Moreover, when details on the nature of links are included, they include quantitative rather than qualitative approaches (focused on assigning numbers to links called weights (Barrat et al. 2004), rather than labels or types). For instance, in the study of mobile phone networks, the frequency and length of interactions has often been used as measures of link weight (Onnela et al. 2007), (Hidalgo and Rodriguez-Sickert 1008), (Miritello et al. 2011). More recently, this literature stream has also begun to focus on multiplex networks, which are networks where nodes have multiple connections among them (i.e. two cities connected by rail and plane), even though the idea of multiplex networks had also been explored by the social science literature (Wasserman and Faust 1994). Still, this has not brought the study of networks by natural scientists closer to the literature advanced in the social sciences, since the focus has been primarily on the generalization of network measures to networks in which multiple links are available and on the mathematical implications for robustness and fragility of networks with multiple links (Mucha et al. 2010), (Bianconi 2013), (Gómez et al. 2013), (Buldyrev et al. 2010).

But the differences between the approaches followed by natural and social scientists do not stop in their conceptualization of what links are, but extend to the link formation mechanisms that they usually use to explain the structure of networks.

Social scientists explain link formation through two families of mechanisms; one that finds it roots in sociology and the other one in economics. The sociological approach assumes that link formation is connected to the characteristics of individuals and their context. Chief examples of the sociological approach include what I will call the *big three* sociological link-formation hypotheses. These are: shared social foci, triadic closure, and homophily.

The social foci hypothesis predicts that links are more likely to form among individuals who, for example, are classmates, co-workers, or go to the same gym (they share a social *foci*). The triadic closure hypothesis predicts that links are more likely to form among individuals that share "friends" or acquaintances. Finally, the homophily hypothesis predicts that links are more likely to form among individuals who share social characteristics, such as tastes, cultural background, or physical appearance (Lazarsfeld and Merton 1954), (McPherson et al. 2001).

The link formation mechanisms favored by economists, on the other hand, favor strategic decisions making. Building on game theory scholars have built (Jackson and Wolinsky 1996), (Jackson 2010) strategic games where self-interested individuals form and severe links as they evaluate the cost and benefits of their interactions. These are network formation mechanisms that are inspired in idea of equilibrium, which dominates neoclassical economics since (Walras 1984) formalized it over a century ago. Yet, strategic games look for equilibrium in the formation and dissolution of ties in the context of the game theory advanced first by (Von Neumann et al. 2007), and later by (Nash 1950).

The link formation mechanisms used by Natural scientists, however, are often not based on strategic games, or dependent on social context, but instead, are based on models that are agnostic about the characteristics of the individuals involved in the formation of a link. For the most part, natural scientists model the evolution of networks as stochastic processes that tie back the evolution of a network back to its structure.

A popular example of such a stochastic model is the idea of preferential attachment, or cumulative advantage. Preferential attachment is the idea that connectivity begets connectivity. More formally, it is the assumption that the probability that a node would acquire a new link depends linearly in the number of nodes that are already connected to it. Preferential attachment is an idea advanced originally by the statisticians John Willis and Udny Yule in (Willis and Yule 1922), but has been rediscovered numerous times during the twentieth century. Willis and Yule were looking to explain the scale-free structure of the networks defined by biological taxonomies (that is, they wanted to explain why some branches in the tree of life branch out much more than others). Yule found that most genera had only one species, but that most species came from a single genus. The explanation Yule gave was that the more species a genus has, the more species it can eventually produce. Rediscoveries of this idea in the twentieth century include the work of (Simon 1955) (who did cite Yule), (Merton 1968), (Price 1976) (who studied citation networks), and (Barabási and Albert 1999), who published the modern reference for this model, which is now widely known as the Barabasi-Albert model.

This growth and preferential attachment model is a perfect example of a network formation mechanism that ties the formation of links to the topology of the network, rather than to individual characteristics of nodes. Preferential attachment, in its pure stochastic interpretation, stands in stark contrast with the models of network formation favored by social scientists because preferential attachment is agnostic about *why* people connect to highly connected nodes, or hubs—it just assumes they do, and then, leverages that assumption to explain a coarse property of the network (its degree distribution). For many social scientists, however, preferential attachment would represent an incomplete explanation of link formation since their main interest would be to understand why people want to connect to hubs. Is it because they have a prestige bias (Henrich 2015)? Are they searching for status? Economic gains? Popularity? Arbitrage Opportunities? For a social scientist, even if all of these alternative hypotheses lead to similar outcomes, separating among them is what it is relevant. In contrast, most natural scientists are happy with a preferential attachment type model since they often consider differences in the reasons why nodes connect to hubs to be irrelevant, especially if these mechanisms do not introduce any changes in the coarse structure of the resulting network. In the language of natural scientists these differences are *symmetries* that give rise to the same universal mechanism: preferential attachment. In the eyes of the social sciences, however, understanding which of all of these hypotheses drives the formation of the network is what one needs to explore.

Another example of a link formation mechanism advanced by natural scientists and that connects the formation of links directly to the topology of the network is the idea *duplication* *and divergence*. In a duplication and divergence model, links are formed as old nodes are duplicated together with a subset of their connections. Think of the biological interactions available to a duplicated protein. If the gene that encodes a protein duplicates, then, the "twin" protein will initially connect to the same proteins than the original protein. Yet over time, one of the two proteins can develop new interactions, and also, lose some of the old interactions it had, since the interactions of a protein are redundant with that of its "twin." As a result, you get a model in which the network grows as nodes are duplicated, and where links grow as these duplicated nodes evolve the set of connections they have. This duplication and divergence models also lead to preferential attachment, since nodes with more links are more likely to see one of their neighbors duplicate.

Duplication and divergence models (Ispolatov et al. 2005), (Vázquez et al. 2002) have been used with great success to explain the structure of biological networks (i.e. protein interaction networks, or metabolic networks), including their heterogeneous degree distribution, modularity (Solé and Fernandez 2003), (Solé and Valverde 2008), and hierarchical structure (Ravasz et al. 2002). Duplication and divergence models, however, are also agnostic about the non-topological characteristics of nodes, and therefore, represent another example of a link formation mechanism that ties the evolution of a network back to its own topology[2].

Juxtaposing the models of link formation advanced by social scientist and natural scientists, however, helps us uncover some important differences between the approaches followed by these two coarsely defined groups of scholars. The link formation mechanisms preferred by social scientists involve a sense of identity and strategy, since they focus on who is connected to whom and why. By contrast, the link formation mechanisms preferred by natural scientists are more neutral, focusing on how connections depend on the position that an individual occupies in a network, but not on who that individual is, or on the strategic choices that pushed an individual to make or cut a connection. As we will see next, these different approaches are justified by different scientific objectives. Stochastic approaches are good at explaining features that are observed over a large variety of networks, what natural scientists call *universal* features, such as the heterogeneous degree distributions of many networks (Albert and Barabási 2002) or their short average path lengths (Watts and Strogatz 1998). When the goal is explaining similarities between networks observed in different systems (from genetic interactions to the physical internet), then it makes sense to use a model that is context agnostic, rather than specific. On the contrary, if a person's goal is to explain and interpret the structure of a narrowly defined network in a specific context, then adopting a context agnostic model will be inadequate, since those models provide answers that are too loosely specified to be informative of the specific social processes driving the network.

In the next section, I continue to explain the differences between the theoretical approaches used to model networks by natural and social scientists by going deeper into the applications used to justify the study of networks. This should help illuminate the preferences for the link formation mechanisms that I have just described.

**Applications of Networks**

Consider the link formation mechanisms that are preferred by sociologists and that we described above as the *big three*: These are homophily, shared foci, and triadic closure. Why would social scientists prefer these link formation mechanisms to stochastic models, such as Yule's preferential attachment process (a.k.a the Barabasi-Albert model)? The answer can be found by asking: what can these link formation mechanisms help explain that Yule's process cannot?

One example is the ethnic and cultural segregation of social networks (Ibarra 1992), (Shrum et al. 1988), (Moody 2001), (Quillian and Campbell 2003), (Currarini et al. 2009). Segregation is a property that is connected to the structure of networks, but that goes beyond it, since it involves the distribution of individual level characteristics, such as the ethnic and cultural background of the individuals in that network. We can explain ethnic and cultural segregation, however, by invoking the *big three* network formation mechanisms of sociology: shared foci, homophily, and triadic closure. Together, these three mechanisms are expected to give rise to homogenous self-reinforcing groups, like the segregated groups we observe in society. Of course, there is more to segregation than what can be explained by these three mechanisms, but this simple example should give you a hint about why they are a better starting point in this case.

As another example consider the labor market, as studied by economic sociologist rather than economists. Economist sociologists, such as Mark Granovetter, have shown that most individuals get jobs from friends and acquaintances (Granovetter 1974). This observation is relevant because it shows that labor markets are *embedded* in social structure (Granovetter 1985) and hence, that the links formed by social mechanisms constrain economic activity (social networks are the "pipes" that determine what economic transactions are possible). This is an observation that also contrasts the theories advanced by new-institutional economists that see social structure as the equilibrium of the institutions that are optimal for a given type of commercial interaction. Yet, Granovetter's empirical labor market results have been reproduced repeatedly and show that social networks drive, on average, roughly half of the labor market (Putnam 2000), (Schwartz 2013). Moreover, Granovetter and others have shown that the jobs assigned through social interactions are primarily the high paid, high-skilled jobs, giving validity to his *embeddedness* theory.

Now, to show how social theories can be combined to advance explanations of complex social phenomenon, let's put together the embeddedness of labor markets and the dynamics of social segregation described above. Together these two mechanisms imply that individuals from different ethnic groups will face different job opportunities (de Souza Briggs 1998). This is another example of a relevant question that is connected to the structure of networks, but that requires a nuanced description of both, the individuals involved in a dyad and of how individual characteristics affect the process of dyad formation.

Labor markets and segregation are two questions that interest social scientists and that require an understanding of networks that goes beyond network topology. Yet, to understand social scientists' description of ties we need to dig deeper and include also their interpretation of ties as the embodiment of trust.

Trust is a dimension of social networks that has been of paramount importance for social scientists, but that has been mostly ignored by natural scientists. The importance of trust in social network literature is well reflected in the literature on social capital. This is a literature advanced by sociologists (Granovetter 1985), (Coleman 1988), (Burt 2001), (Burt 2005), political scientists (Fukuyama 1995), (Putnam 2000), and economists (Alesina and La Ferrara 2002).

Social networks and trust are intimately connected, since individuals are more likely to trust those with whom they share social connections, interact frequently, and share friends and acquaintances with (Granovetter 1985), (Burt 2001), (Burt 2005), (Coleman 1988). Yet, not all social connections embody trust. Trust, however, also helps us interpret the emergence of triadic closure, since people connect to friends of friends because they are more likely to trust them—you can think of the connection to the mutual friend as a form of insurance (when people consider connections to be valuable). Going back to our labor market discussion, trust can also be used to explain the role of social networks in the labor market, since the willingness of people to hire friends of friends could be seen as a reflection of the trust that flows indirectly through an open triad—or of the insurance represented by the mutual friend. Of course, friends are also likely to have similar skills, so homophily is expected to reinforce Granovetter's labor market results. Finally, trust can also be used to explain the size of the firms that populate an economy. As Francis Fukuyama argued in his book *Trust* (Fukuyama 1995), economies where people are more likely to trust strangers will form larger social and professional networks and will gravitate towards complex industries (such as aircraft manufacturing)[3]. Finally it is worth noting that trust, through the theory of social capital, has been connected with long-term economic growth—even though these results are based on regressions using extremely sparse datasets. Nevertheless, the evidence suggests that social capital and social institutions are significant predictors of economic growth, after controlling for the effects of human capital and initial levels of income (Knack and Keefer 1997), (Knack 2002)[4]. So trust is a relevant dimension of social interactions that has been connected to individual dyads, network formation, labor markets, and even economic growth.

People studying trust have also been able to connect trust to other social institutions, such as the family. In fact, societies where individuals rely more heavily on family links are also societies where individuals are less likely to trust strangers, and consequently, less likely to engage in political and civic participation (Fukuyama 1995), (Alesina and Giuliano 2011). Moreover, some of the social and economic correlates of family relationships are known to survive in the families of immigrants, suggesting that the effect of social institutions in the type of links that a society forms is long lasting (Alesina and Giuliano 2010)[5].

So what are the applications that interest natural scientists? Natural scientists have not focused primarily on trust, labor markets, or social segregation. Instead, they have focused mainly on five things: (i) explaining the topology of networks in terms of stochastic models, (ii) developing algorithms to quantitatively describe the topology of networks, from their degree distribution to their community structure, (iii) modeling the spread of diseases and information on networks, (iv) using networks as a mean to model large interconnected systems, by mapping connections among diseases, language, or similar products, and (v) to study the implications of network structure for game theoretical outcomes, not in the context of link formation, but primarily in the context of the evolution of cooperation. Goals (iii) and (v) are shared among natural scientists and social scientists, in part, because the puzzle of cooperation is one of long tradition in both evolutionary biology (Dawkins 1976) and economics (Von Neumann et al. 2007).

The first two goals of natural scientists, explaining network formation through stochastic models and quantifying network structure are highly intertwined, since natural scientists use the structural features of networks to validate the predictions of their stochastic models. This has lead natural scientists to create a vast literature on the empirical characterization of network structure which focuses on looking at a network's degree distributions (Barabási and Albert 1999), (Albert and Barabási 2002), (Krapivsky and Redner 2001), its hierarchical structure (Trusina et al. 2004), (Ravasz et al. 2002), (Clauset et al. 2008), community structure (Palla et al. 2005), (Ahn et al. 2010), (Girvan and Newman 2002), (Blondel et al. 2008), (Fortunato 2010), and also, the likelihood of hubs to connect to hubs. This last property is usually studied under the name of degree-degree correlations (Newman 2002), but alternative ways of measuring this property (which focus on clustering among hubs) have been rebranded, as the "rich-club" phenomenon (Colizza et al. 2006), or "fractal" networks (which are networks in which hubs have a tendency to repeal each other) (Song et al. 2005), (Song et al. 2006). The tendency for hubs to connect to hubs, however, is an idea that is closely related to homophily, but in the hands of natural scientists shows their preference for topological measures, since degree correlations study the tendency for links to form among pairs of nodes characterized by a topological feature (their degree) instead of an intrinsic property (such as their ethnic group, or gender), or and acquired property (a node's income, or social status).

But what questions can natural scientists answer with their context agnostic approaches? Some questions that are popular among natural scientists are questions of percolation, in which the vulnerability of networks to the removal of nodes due to errors and attacks is studied (Albert et al. 2000), (Cohen et al. 2002), (Achlioptas et al. 2009). Also, topological approaches are popular in the link prediction literature, which is popular among computer scientists and has applications for social media companies (which are trying to predict new friendships, followers, or clicks on ads., all of which are new links) (Liben-Nowell and Kleinberg 2007), (Clauset et al. 2008), (Lichtenwalter et al. 2010).

Yet the link prediction literature is a good example of a disconnection between the literatures advanced by natural and social scientists. Even though all link prediction papers build heavily on measures of triadic closure (Liben-Nowell and Kleinberg 2007), (Lichtenwalter et al. 2010), they often do not cite the social science literature on triadic closure. Instead, they focus on comparing a repertoire of measures of open triads and machine learning algorithms in search for the combination of features and algorithms that maximize the accuracy of the predictions.

> One place where natural scientists have been relatively successful at is at using the idea of a network to map connections in non-social systems. This usually involve taking a bi-partite network, like the network connecting diseases to genes (Goh et al. 2007), countries to products (Kramer et al. 2014), or languages to people (Ronen et al. 2014), and creating a projection to connect diseases that share genes, products that are exported in tandem, or languages that are co-spoken. In the context of medicine and biology these ideas are manifested in the new literature on network medicine, which is based on the creation of networks connecting diseases that are caused by the same genes (Goh et al. 2007), that share metabolic paths (Lee et al.

2008), or that affect the same patients (Hidalgo et al. 2009) These disease networks are being used to identify new disease genes, and uncover the biological significance of disease-associated mutations (Barabási et al. 2011).

But there are also applications of networks that interest both natural and social scientists. One of these is the spreading of epidemics and information. The basic question that this literature tries to answer is where people get new information from (for instance, about a job), or how diseases spread. Of course, the position that a person occupies in a network should affect the information that is available to him or her, or the probability that a person interacts with another individual that is carrying a disease.

In the context of the natural sciences this literature has emphasized the development of mathematical models of disease contagions. Following the pioneering work of William Kermack and Anderson McKendrick (Kermack and McKendrick 1927; Kermack and McKendrick 1932; Kermack and McKendrick 1933), many scholars have explored the consequences of incorporating networks structure explicitly in the process of epidemic spreading (Pastor-Satorras and Vespignani 2001), (Barrat et al. 2008), (Boguñá and Pastor-Satorras 2002), (Colizza et al. 2006a), and also, of including other effects, such as differential susceptibility—the fact that not all nodes are equally vulnerable to a disease (Smilkov et al. 2014)—into these models. Social scientist, on the other hand, have focused on what ties are more likely to bring in new information, which are primarily weak ties (Granovetter 1973), and on why weak ties bring new information (because they bridge structural holes (Burt 2001), (Burt 2005)).

In recent years, the studies of diffusion processes in networks have been expanded to works that extends beyond the spreading of infectious diseases, or information about jobs, and now include the spread of behaviors and health conditions, such as obesity and smoking (Christakis and Fowler 2007), (Christakis and Fowler 2008), the diffusion of innovations (Rogers 2003), behaviors (Centola and Macy 2007) (Centola 2010), emotions (Kramer et al. 2014), and even the industrial structure of economies (Hidalgo et al. 2007).

Finally, we have the literature connecting game theory and networks in the context of the evolution of cooperation. The evolution of cooperation is a classic scientific question since there are many situations where individuals have an incentive to cheat, making the prevalence of cooperation a deep theoretical puzzle. The original attempts to explain the emergence of cooperation in large populations focused on the role of strategies involving punishment schemes and reciprocity (Axelrod and Hamilton 1981). More recently, however, heterogeneous networks have been found to be effective promoters of the evolution of cooperation, since there are advantages to being a cooperator when you are a hub, and hubs tend to stabilize networks in equilibriums where levels of cooperation are high (Ohtsuki et al. 2006), (Pacheco et al. 2006), (Lieberman et al. 2005), (Santos and Pacheco 2005). These results, however, have also been challenged by human experiments finding no such effect (Gracia-Lázaro et al. 2012). The study of cooperation in networks has also been performed in dynamic settings, where individuals are allowed to cut ties (Wang et al. 2012), promoting cooperation, and are faced with different levels of knowledge about the reputation of peers in their network (Gallo and Yan 2015). Moreover, cooperating behavior has seen to spread when people change the networks where they participate in (Fowler and Christakis 2010).

**Building bridges**

In the last sections I juxtaposed the literature of social scientists and natural scientists working on networks, two groups of academics that often fail to see eye to eye. This juxtaposition helped us illustrate important differences between the methodology and questions explored by each of these groups of scholars. Scholars trained in the social sciences focus on explaining social and economic phenomena, and are interested on how networks affect the individuals and organizations forming these networks (demographics, income, etc.). As the sociologist Linton Freeman remarked in *The Development of Social Network Analysis*:

> *"The social network approach is grounded in the intuitive notion that the patterning of social ties in which actors are embedded has important consequences for those actors. Network analysts, then, seek to uncover various kinds of patterns. And they try to* *determine the conditions under which those patterns arise and to discover their consequences."* (Freeman 2004)

Natural scientists, on the other hand, are interested in identifying features that are common to a wide variety of networks, and hence focus on the use of stochastic and generative models that are agnostic about the properties of individuals, or their goals. This pushes natural scientists to focus on what different networks have in common, instead of what sets them apart. As Barabási explains in Linked(Barabasi 2014):

*"The diversity of networks in business and the economy is mindboggling. There are policy networks, ownership networks, collaboration networks, organizational networks, network marketing-you name it. It would be impossible to integrate these diverse interactions into a single all-encompassing web. Yet no matter what organizational level we look at, the same robust and universal laws that govern nature's webs seem to greet us."*

Yet, despite their difference in focus, each literature has been able to make great advances. While social scientists have made great progress in questions that need to be understood in a nuanced social context, like the role of trust on labor markets, natural scientists have advanced the understanding of network questions that are not context specific, and are governed by general constraints. But can these approaches learn from each other?

Both of these approaches can benefit from each other, since natural scientists often throw the baby with the bathwater when exploring social questions in absence of a well defined social context, or by not considering the multiple hypotheses that a social context can imply. On the other hand, social scientists often have problems seeing explanations that are based on statistical properties or constraints that are independent of context, since they have developed a strong taste for theories that are more teleological than those advanced by natural scientists. So they can see mirages of mechanisms in situations where an explanation based on constraints is enough.

And in this taste for teleology is where we find one of the great differences between social scientists and natural scientists, since these differences bring each of these disciplines to a different interpretation of what they mean by answering the question: "*why*?" Social scientists look for answers to why questions that involves the purposeful action of actors, no matter whether those purposes are driven by self-interest (like in the economics tradition), by a process of socialization (like in sociology), or whether they developed in a struggle for power (like in political science). Natural scientists, on the other hand, answer why questions by looking at the constraints that limit the behavior of the system. This is an approach that builds on the tradition of physics, since the earth does not orbit the sun[6] for a purpose, but because the law of gravity acts as a constraint (metaphorically, as a tense rope) that shapes our planet's elliptical motion. By the same token, the reason *why* momentum is conserved in many physical systems is because the Hamiltonian of these systems (the Hamiltonian being an advanced way of representing the constraints of a dynamical system) does not depend on that systems' position. Why questions do not always involve purpose, but it is important to note when they do.

So can these literatures come together?

I think there are two ways in which they can. One is by creating teams that use the diversity of skills found in scholars from different disciplines as an advantage. The other one is to focus on topics that are of common interest to scholars from the social and natural sciences, such as online social interactions.

So let's look at the first of these two options. Scholars from the natural and social sciences have a diversity of skills that when put together can be very powerful. Social scientists are often great narrative theoreticians, and are great at framing arguments and highlighting the social relevance of findings. Also, social scientists are trained to think in terms of multiple chains of causations, so they are good at identifying potential underlying assumptions and hypotheses. They also have a good toolbox of quantitative techniques they can use to separate among multiple hypothesis, from simple multivariate regressions, to matching methods, and instrumental variables. Natural scientists on the other hand, are comparatively skilled in the development and implementation of new algorithms and metrics, and are often better at the use of graphical statistical methods, which in presence of the right renormalization techniques can help uncover universal distributions. Also, natural scientists have a natural tendency to think of statistical controls in terms of null models. In network science, these null models are useful because they help discount patterns that are explained by simple structural features, like a network's degree distribution (Maslov and Sneppen 2002), (Vázquez et al. 2004), (Hidalgo and Hausmann 2009). Moreover, computer scientists tend to be good at optimizing algorithms, which is something required for scaling research to large datasets. So in principle, collaborations between social scientists and natural scientists could result in high quality work because natural and social scientists have a larger and more powerful toolbox when working together than in isolation.

In fact, there are quite a few examples of successful work involving collaborations between natural and social scientists. These involve the work by the sociologist Matthew Salganik, and the physicists Peter Dodds and Duncan Watts in market forces' (Salganik and Watts 2008), or the work by the sociologist Brian Uzzi, the economist Benjamin Jones, and the physicist Stephan Wuchty on knowledge production by teams (Wuchty et al. 2007), (Jones et al. 2008). Other examples include the collaborations between the physicist Cesar Hidalgo and the economist Ricardo Hausmann in economic complexity and economic growth (Hidalgo et al. 2007), (Hidalgo and Hausmann 2009).

The other way in which these two literatures can come together is less methodological and more topic-oriented. In fact, there are many topics that are of the interest of both natural and social scientists. Two that I mentioned previously are the diffusion of information and contagious diseases, and the evolution of cooperation. Another topic, of more recent appearance, is social media and its effects in society. In recent decades sociologists, like Barry Wellman, have written extensively about how modern communication technologies are affecting social structure. In the early 2000s Wellman (Wellman 2001) begun countering Robert Putnam's claim that social capital was declining (Putnam 2000), and argued instead that social capital was moving online (Wellman et al. 2001). More recently, Wellman and Lee Rainie summarized this argument in the idea of networked individualism, the idea that individuals are no longer bound to closely-knit groups, but are instead nodes in sparser global networks (Rainie and Wellman 2012).

But natural scientists are also interested in online social networks, and they have been good at developing scalable algorithms to help analyze large samples of these networks. Their focus has been on identifying influential individuals in social media (Cha et al. 2010), (Bakshy et al. 2011), verifying the veracity of information (Castillo et al. 2011), and performing sentiment analysis (Bollen et al. 2011), (Go et al.), (Dodds et al. 2011). So online social behavior could be a new opportunity for these literatures to come together.

But beyond topics, and skills, there are still some important differences in the format and style of publications that can limit cross-collaboration among scholars working on different parts of the fragmented network literature.

One of these formal aspects is the enormous difference in the formats of publications that are preferred and accepted in the natural and social sciences. Differences in format may seem cosmetic, but due to the social (or antisocial) nature of peer-review, differences in the expectations that academics have with respect to format can result in papers being quickly misunderstood, and rejected, by scholars trained in different fields.

One important difference here is the role of an introduction in a paper. In the natural sciences, especially in physics, introductions are considered boilerplate summaries of previous research that are mostly irrelevant, since what makes or break a paper is the results section. That is why in the natural science literature there are so many papers that start with a variant of the generic sentence: *"In recent years there has been much interest in the study of networks."* In the social sciences, however, the introduction is essential to the paper, since it is the place where scholars fully explains his or her contribution in the context of what is known. These differences also translate into the length of the papers. Natural science papers tend to be extremely short in length by social science standards (usually less than 4,000 words), and hence, economize language in their introductions and literature reviews (If this was a natural science paper, it would have ended more than 4,000 words ago). Often, natural scientists cite literature in one or two short paragraphs, instead of dedicating a multipage section detailing the contributions of other scientists. Social scientists on the other hand, write extensive literature reviews in which many of the papers cited are described in multiple paragraphs. Social science papers put substantial effort on discussing the previous literature before presenting any of their own work, and are often rejected if they fail to provide a good review of the literature. Unfortunately, these styles are incompatible. Write a natural science introduction for a social science audience and your paper will be rejected before the reviewer sees the results section. Write a social science introduction for a natural science audience and you will be scoffed away for being "unnecessarily verbose."

Another formal difference involves the use of graphical statistical methods (Hidalgo 2010) and multivariate statistics. The first ones are preferred by natural scientists and often avoided by social scientists, while the reverse is true for multivariate statistics. These differences, however, are also misinterpreted as shortcomings since social scientists often think of graphical statistical methods as "non-serious," since they are limited in their ability to control for co-founding factors, while natural scientists find that the use of tables, instead of graphical representation of results, occludes information about functional forms, which natural scientists consider important.

In this paper I provide a brief and incomplete review of what is a large and fragmented literature on network science. Hopefully, the juxtaposition presented here helps explain the value of the approaches followed by academics in both of these streams of literature and helps stimulate further discussion in the study of systems of organized complexity.

Wuchty S, Jones BF, Uzzi B (2007) The Increasing Dominance of Teams in Production of Knowledge. Science 316:1036–1039. doi: 10.1126/science.1136099

---

[1] For instance, in this review I will not include the life sciences among the natural sciences even though they have done extensive work on biological networks. My decision not to include the work of biologists in this review is to simplify the scope. Also, I will not discuss financial networks, or graph theory. For the most part, I will be discussing work involving social networks (networks where nodes are people), as these networks have been of the interest of both social scientists and natural scientists.

[2] Certainly, saying that natural science approaches focus only on link formation mechanisms that tie back to topological features is a tad unfair. After all, this is more a matter of emphasis than an absolute claim. In fact, in the natural science literature there are approaches to link formation that do focus on the non-topological characteristics of nodes. A good example here is the introduction of the idea of a node's *fitness* (Bianconi and Barabási 2001a)(Bianconi and Barabási 2001b). A node's *fitness* is an exogenous parameter that models the attractiveness of linking to a node, and that was introduced to destroy the strong correlation between a node's age and connectivity that is implied in a model based purely on growth and preferential attachment (Adamic et al. 2000). The treatment of fitness in the natural science literature, however, has been mostly abstract. Fitness has a distribution and a value, but not a unique or even narrow interpretation in terms of a social or economic characteristic. Moreover, little effort has been made to link fitness to one of its many possible interpretations. In high-school friendship networks, is fitness a reflection of the physical beauty of a student or its sport prowess? In networks of commercial interactions, is fitness a reflection of the quality of service, or the marketing muscle of a firm? These questions are here to illustrate the contrast between the interests of natural and social scientists, since these questions would be more of the interest of social scientists than of natural scientists. The latter would be mostly content with assuming that differences in fitness affect the evolution of network structure, while the former would want to know why some individuals are more attractive than others, even if these reasons do not change the overall structure of the network.

[3] Here it is worth noting that there are important cases where these large efforts in socialization emerge as a consequence of state intervention, like the aircraft industry in Brazil, or in France, as described by Fukuyama. In any case, the spontaneous emergence of large networks in societies endowed with trust tends to be a more successful and rapid form of economic development than the one that is forced by state interventions, which have a low success rate.

[4] It is worth noting that these empirical results hinge on small sample sizes, since data on trust is available for a few countries over relatively short time periods.

[5] At the individual level, low trust is associated with traumatic experiences, belonging to a group that historically felt discriminated against, being economically unsuccessful in terms

of income and education, and living in a racially mixed community or one that is unequal in terms of income and education (Bianconi and Barabási 2001).
[6] or more precisely, the center of mass between the earth and the sun.